\documentclass[aps,twocolumn,showpacs,nofootinbib]{revtex4}
\usepackage{graphicx}
\newcommand{\beq}{\begin{equation}}
\newcommand{\eeq}{\end{equation}}
\newcommand{\beqa}{\begin{eqnarray}}
\newcommand{\eeqa}{\end{eqnarray}}

\def\half{\frac{1}{2}}

\def\opone{\leavevmode\hbox{\small1\normalsize\kern-.33em1}}

\begin{document}

\title{Is realism compatible with true randomness\footnote{\copyright ~Oxford University 2010.}?}

\author{Nicolas Gisin \\
\it \small Group of Applied Physics, University of Geneva, 1211 Geneva 4,    Switzerland}

\date{\small \today}

\begin{abstract}
It is argued that realism and true randomness are fully compatible. Realistic true random events are acts of pure creation that obey strict laws, but do not necessarily satisfy Kolmogorov's axioms of probabilities. Realistic true randomness is some sort of nondeterministic force, or propensity of physical systems to manifest such and such properties under such and such conditions. Realistic random events reflect preexisting properties, as required by realism, simply the reflection is not deterministic; still, the preexisting properties determine the propensities of the different possible events.

It is argued that deterministic extensions of quantum physics are necessarily incompatible with special relativity. Hence, from today's violations of Bell's inequalities one can conclude that all future physics theories will display true randomness as does quantum physics.

It is argued that accepting true randomness and realism leads to new questions with interesting answers, allowing one 1) to study nonlocality in configurations with many independent sources and 2) to bound how much free will is needed for a proper violation of Bell's inequality.
\end{abstract}

\maketitle

\section{Introduction}\label{intro}
The possibility of elaborating connections between fundamental concepts like {\it realism}, {\it locality} or {\it free will} on the one hand, and hard sciences like physics or biology on the other is fascinating. But it comes along with the danger of vagueness, if not blatant vacuous phraseology (especially when physicists express their views). Hence, I won't attempt to (re)define realism or free will, but rather try to illustrate my purpose with what I consider insightful connections to parts of physics that I am familiar with. I'll try to follow the example of {\it locality of nature} and Bell inequalities, an example that Abner Shimony named experimental metaphysics \cite{ExpMeta}.

Recently we learned that connecting a fundamental concept like locality to quantum physics, thanks to the work of Bell and many others, allows one to investigate new applications. I consider this as the signature that the connection is deep and not merely superficial: from deep connections new questions, and also new applications ought to emerge. For example, the connection between {\it locality of nature} and Bell inequalities has gone way beyond what Bell and other precursors had originally in mind. Today, Bell inequalities cross-fertilized with Quantum Information Science has led to the concept of {\it Device-Independent Quantum Key Distribution} (DI-QKD) \cite{QKD-DI-PRL,QKD-DIFullLength}. This is a protocol for establishing a cryptographic key between two distant partners in which the security is not based on Hilbert-space quantum mechanics, as it is for standard QKD, but the security is based directly on the violation of a Bell inequality, that is on quantum nonlocality \cite{bhk,Acin06}. The intuition is very simple: if no local variable can describe the correlation, then no adversary can possibly hold a copy of these non existing local variables. Despite the simplicity of the intuition, it took a long time to make it precise and it is still an active field of research with plenty of open questions. Such a connection also gives a welcome new push towards detection loophole free experiments (especially with photons over tens of kilometers). Suddenly the infamous detection loophole is part of applied physics \cite{PearleDetLoophole,DIQKDepxProposal}!

This note is organized as follows. First, in section \ref{realism}, I'll ask what is realism for a quantum physicist. I'll argue that there is no contradiction between realism and non-determinism: a non-deterministic world can be as real as a deterministic one. Actually, our world is non-deterministic and real. Next, in section \ref{locality}, building on the previous one I ask what additional variable in a non-deterministic theory, like quantum physics, could mean. I'll illustrate my point by showing how different concrete answers lead to different new results. Finally, in section \ref{freeWill}, I admit true randomness to illustrate the new questions and applications that follow; surprisingly these deal with quantifying free will \cite{Popescu}.

\section{Realism and true randomness}\label{realism}
Realism is often "defined" as stating that there is something out there and that we can interact with it. This is too vague for direct usage in physics. At least one should be precise about what it means to interact with reality. Must the interaction be a direct one? or could a mere indirect interaction suffice? and what is meant by an indirect interaction? Furthermore, should the interaction be 2-way?

Often (too often) physicists reduce realism to the idea that each physical quantities always possesses a value. But then, either this value is unaccessible, hence unphysical. Or this value can be revealed by appropriate measurements (to arbitrary good approximation, at least in principle). But then, these measurements have predetermined outcomes: the world is deterministic and realism is nothing but a fancy word for determinism.

Let us consider a world in which some measurements produce truly random results. In such a world, some events are fundamentally not predetermined; not only do we, humans, not know them in advance, and there is no way for us to know them in advance, but even nature doesn't know them in advance. Why should such a world not be real? We might be influenced by our cultural context: God has all power, including that of knowing the future, hence the future can't be really open. Fine, we are certainly influenced by our culture, but this is certainly not an admissible argument for scientists. So, what opposes true randomness and realism? Maybe the fear that true randomness implies true becoming, i.e. spontaneous acts of creation? Yes, but why should acts of creation (non-predetermined events that just happen) not be real? Is it the fear of chaos? Such a fear seems inappropriate in a period governed by quantum physics: we know that randomness can be very well organized and structured by strict laws. Finally, the problem might be that a random event seems to emerge from outside the world: since before its realization it was nowhere inside the world, it had to come from outside.

I am afraid that we have now reached a point of sufficient vagueness for all possible answers to be equally possible/impossible. Let's be pragmatic: a pure random event just happens, it follows some probability laws (i.e. has a well defined propensity to manifest itself), but it comes from nowhere. And this doesn't make it any less real than deterministic events like the arrival of a train in a station.

What determines the probability (propensity) of measurement results? The answer is that {\bf measurement results reflect preexisting properties: this is realism!} Simply, the reflection is not a deterministic one, the preexisting properties only determine the propensities\footnote{The word probability is so much linked in physicists mind to Kolmogorov axioms that one should avoid it. Indeed, probabilities satisfying Kolmogorov axioms can always be interpreted as epistemic and one is thus tempted to add hitherto "hidden variables". Hence it is advantageous to use a different word like propensity. This concept is closed to Poper's \cite{PoperPropensity} (whose poor understanding of the EPR argument obscured his deep plea in favor of propensities). In \cite{GisinPropensity} I argued that the propensities in quantum physics are minimal generalizations of determinism in the sense that the Hilbert space structure is such that for all pure states the set of elements of reality (i.e. physical quantities that posses a deterministic value) uniquely determine the propensities of all physical quantities.}, i.e. the natural tendencies, of the different measurement results. Note that this is absolutely compatible with standard quantum mechanics: the probabilities of measurement results are defined by the reduced density matrix. Hence, randomness is not a problem for realism.

Leggett suggested that the preexisting properties should be pure quantum states and that these pure states define the local probabilities in the usual quantum way \cite{Leggett}. This model has been experimentally falsified \cite{Leggettexp}. Surprisingly some physicist insist that realism implies that the preexisting properties are pure quantum states and from the falsification of Leggett's model conclude that realism is falsified. But, clearly, only their limited concept of realism (local pure states with arbitrary nonlocal correlations) has been falsified!

There remains the question concerning when random events happen. I don't know any better answer than the one proposed by the French philosopher Cournot \cite{Cournot}: random events happen at the meeting point of two causal chains, as - I like to add - at the meeting point of a quantum causal chain (described by Schr\"odinger's equation) and a classical measurement apparatus. Admittedly I don't know how to characterize classical measurement apparatuses. Possibly one has to go all the way and consider the second causal chain in Cournot's argument as triggered by free will? Alternatively, one has to assume that all evolutions are stochastic (i.e. the Schr\"odinger equation is only an - excellent - approximation), as in GRW-like models \cite{GRW,BellGRW,Gisin89,PercivalPSD}.

Accordingly, true randomness is compatible with realism. But, some may further argue, determinism is nicer. So why should one believe in true randomness in physics? After all the de Broglie-Bohm pilot wave model shows that it is not necessary. This claim, however, is wrong, as shown in \cite{GisinNoCovariantVariable} and emphasized in section \ref{locality}: quantum physics and special relativity exclude any deterministic extension of quantum physics (but admittedly one may assume that relativity is incomplete \cite{GisinRelativityIncomplete}, e.g. that there is a universal privileged reference frame and that a de Broglie-Bohm model describes reality in this privileged frame, see however \cite{SalartNature,Cocciaro10}.

In conclusion, although some questions remain open, I see no reason to doubt that true randomness is compatible with realism.

\section{What is local in local variables?}\label{locality}
In this section I elaborate on the physical meaning of $\lambda$, the historically so called {\it local hidden variable}, named shared randomness by today's computer scientists, and whose reality is denied by some who hope to thus get rid of nonlocality.

In any derivation of a Bell inequality one always encounters the following factorization condition, also called separability condition:
\beq \label{BellLocality}
p(a,b|x,y,\lambda)=p(a|x,\lambda)\cdot p(b|y,\lambda)
\eeq
where $a$ and $b$ are the measurement results (the outcomes) secured by the two usual players Alice and Bob when they perform the measurements $x$ and $y$ (the inputs).

But what is $\lambda$?

The usual answer runs as follows. $\lambda$ denotes a well localized beable attached to particles \cite{BellSpeakable}. According to this picture, the local $\lambda$ is produced by the same source as that emitting the entangled particles and copies of $\lambda$ are carried by each of the particles. Hence, if the particles are widely separated in space, then the outcome probability at Alice side can depend only on the local measurement setting $x$ (Alice's input) and the local beable $\lambda$, and similarly on Bob's side. The factorization condition (\ref{BellLocality}) is thus a consequence of locality:
\beqa
\lambda&=& local~ beable~ \&~ locality \\
&\Rightarrow& factorization~ condition~ (\ref{BellLocality}) \\
&\Rightarrow& Bell~ inequality
\eeqa
Note that there is no need to assume that $\lambda$ determines the measurement outcomes, it suffice that $\lambda$ determines the outcome propensities. Hence, denying determinism is clearly not sufficient to avoid Bell inequalities and their consequences, i.e. quantum nonlocality.

The above idea is the standard one; it is also the one John Bell had in mind when he derived his famous inequality (recall that Bell was greatly influenced by Bohm's pilot-wave model in which $\lambda$ denotes the particles's position) \cite{BellSpeakable}. But, let's go further. If this picture tells us something deep, then there ought to be consequences beyond Bell inequalities. And indeed there are. If 2 independent sources produce independent pairs of entangled particles, as in entanglement swapping \cite{EntSwap}, then there should be 2 independent sets of local variables $\lambda_1$ and $\lambda_2$.
This is interesting and does indeed lead to new tests of this kind of local variables. In \cite{bilocality} we proved that a visibility higher than $\half$ in entanglement swapping experiments between independent sources, as e.g. \cite{Matthaeus}, suffices to falsify such models. This is significantly easier than violating the usual CHSH-Bell inequality that requires a visibility $>\sqrt{1/2}$. It is surprising and disappointing that it took more than 60 years from Bell's breakthrough to indepth study of the consequences of treating $\lambda$ as local beable. This illustrates the damage when realism is not considered seriously.

But there is a quite different way of looking at the $\lambda$ in the factorization condition (\ref{BellLocality}). Let us first consider standard quantum physics. In this case, $\lambda$ stands for the usual quantum state $\Psi_{AB}$ and condition (\ref{BellLocality}) is nothing but the condition that the quantum state is separable. Next, consider a future theory, a theory that surpasses quantum theory. At present, we do not know this theory, but we known that any physical theory makes predictions. Hence, $\lambda$ could equally well stand for the physical state of the two systems in Alice and Bob's hands as described by any possible future physics theory, i.e. $\lambda$ determines $p(a,b|x,y,\lambda)$. As we do not know this theory, we do not know how to prepare a given $\lambda$ state, but we know that the preparation we carry out today in our labs correspond to a mixture of the $\lambda$'s of this future theory. Hence, condition (\ref{BellLocality}) is an assumption about any possible future theory. The fact that Bell inequalities are experimentally violated today implies thus that any possible future theory (in agreement with today's experiments) displays nonlocality.

Actually, one can go even further and interpret $\lambda$ as the state of the entire universe (as described by today's quantum theory or as described by any possible future theory). In such a view $\lambda$ is not localized. But one can nevertheless study the consequences of the locality (or separability) assumption (\ref{BellLocality}). The only assumption we need in order to make sense of (\ref{BellLocality}) and of the Bell inequalities is that the two inputs, $x$ on Alice and $y$ on Bob's side, are independent of $\lambda$. Hence $\lambda$ denotes the state of the entire universe, except the inputs $x$ and $y$, i.e. Alice and Bob enjoy free will even given $\lambda$ (or there are random number generators independent of $\lambda$): $p(x,y|\lambda)=p(x,y)$ for all $\lambda$.

Notice the difference. Historically $\lambda$ was considered only as a local beable, but actually one can consider $\lambda$ as the real state of the entire universe, except only the two inputs $x$ and $y$. That is, $\lambda$ does not at all need to be local: {\bf $\lambda$ can be considered as a nonlocal beable}, it suffices to assume that the outcome propensities of the local measurements are determined by the local inputs and the global state $\lambda$.

Hence, the violation of a Bell inequality not only means that quantum physics is nonlocal (predicts nonlocal correlations), but that any future physical theory is likewise nonlocal.

Again, such a view of $\lambda$ as the physical state of the entire universe (except $x$ and $y$) as described by any future theory ought to have consequences. Let's thus consider such nonlocal beable $\lambda$ (i.e. without the assumption (\ref{BellLocality})), but assuming them covariant in the sense of special relativity. Because I do not know how to describe true probabilities in special relativity, I'll restrict the analysis to deterministic covariant nonlocal $\lambda$'s. If Alice makes her measurement first, in some reference frame, then her outcome $a$ is a function of $\lambda$ and of her input: $a=F_{AB}(x,\lambda)$, and Bob's outcome is a function of his input $y$, of $\lambda$ and possibly of Alice's input $x$ (because Bob makes his measurement after Alice): $b=S_{AB}(x,y,\lambda)$. In another reference frame in which Bob makes his measurement first, the situation is symmetric: $b=F_{BA}(y,\lambda)$ and $a=S_{BA}(x,y,\lambda)$, where the subscript $BA$ reminds us of the time ordering. If $\lambda$ is covariant (i.e., more precisely, the outcomes determined by $\lambda$ are covariant), then the outcomes should be independent of the reference frame: $a=F_{AB}(x,\lambda)=S_{BA}(x,y,\lambda)$. Hence the function $S_{BA}(x,y,\lambda)$ has to be independent of $y$. But then the pair of functions $F_{BA}(y,\lambda)$ and $S_{BA}(x,\lambda)$ define a local model in the sense of (\ref{BellLocality}), which contradicts well confirmed quantum predictions.  This simple argument was seemingly first dispelled in \cite{GisinNoCovariantVariable}. Hence, studying nonlocal $\lambda$'s tells us something important: they can't be deterministic (see also \cite{FWTrGRWf,RennerColbeclExtention}). Note that as a consequence, they neither could be probabilistic in the usual sense of probability (i.e. satisfying Kolmogorov's axioms), since usual (Kolmogorov) probabilities can always be interpreted as epistemic probabilities: by adding a random variable to $\lambda$ the theory becomes deterministic.

In summary, true randomness will be part of all future physics theory.

\section{Can Free Will be Quantified?}\label{freeWill}
In a deterministic world there is no free will. At best there could be the illusion of free will. But there is nothing I know deeper than the fact that I enjoy free will. How could science question this conviction? In order to do science one has to be able to test theoretical models. But one can perform experimental tests only if one is able to freely chose which tests to perform and when to perform them. Hence, without free will science is impossible.
Consequently, science can't even question the existence of free will\footnote{Note that our experience of free will comes first, also with respect to philosophy. Hence, no rational argument against free will can be relevant.}. Or science would also be an illusion: we could live in a world ruled by some laws, but be determined (programmed, as in the film Matrix) to perform those experiments that delude us to believe in totally different laws.

{\bf Hence, any serious scientist has to believe in free will.}

Determinism is incompatible with free will, but randomness is clearly not sufficient to explain free will. Free will, by essence, seems to emerge from outside space-time, i.e. it is not part of physics. Nevertheless, accepting free will and true randomness in a realistic world-view allows one to ask new questions and work out new applications.

First the application. If quantum measurement results are truly random, then they provide us with truly random numbers. Hence, Quantum Random Number Generator (QRNG) is a natural application. QRNG is actually already a commercial success: one can buy little devices of the size of a match box that deliver megabits of quantum random numbers per second \cite{idQ}. But for the purpose of this note, this might be too simple: the outcomes of the QRNG could actually have been produced much earlier, memorized in the "match box" and delivered on demand. But consider the following. The outcomes of measurements in Bell tests are random and their correlation guarantees that they were not produced in advance, at least not before the inputs were given. Hence, if the inputs are random and the correlation violate some Bell inequality, then the outcomes are necessarily also random and unique. Now, the input bit string can be shorter than the output, i.e. one consumes less input random bits than the number of output random bits produced. For example the input could be binary and the outcome ternary. This offers a new application: randomness expansion \cite{RandomnessExpansionCollbeck,RandomnessExpansionPironio}! One can even go a step further. The input bits could be freely chosen by humans. The input bit rate would be very low, at best one freely chosen bit per second or per minute. But using the Bell-based randomness expansion several times in series one could get arbitrarily high rates of truly random bits, a very valuable resource for our information based society.

Next, the new question. Free will is needed for science. But how much free will is required? More precisely, assume we do not enjoy full free will, then how much {\it lack of free will} is still compatible with science? Let me illustrate these seemingly impossible questions, we'll see that one can quantify the maximum lack of free will compatible with proper tests of Bell's inequality.

Let me first explain what I mean by {\it lack of free will}. Assume one has to make a choice between $N$ possibilities. Hence, one needs $\log_2(N)$ bits of free will. Now, assume that actually we are partially determined and that in fact we can only chose between $M$ possibilities ($M<N$): although we have the illusion to chose between all $N$ possibilities, we actually chose only between M possibilities, i.e. we are programmed not to chose any of the remaining $N-M$ remaining ones. Then our {\it lack of free will} would be $\log_2(N)-log_2(M)=\log_2(N/M)$. For example, if one has to chose between 4 possibilities, $c_1,c_2,c_3,c_4$, but if one is actually determined to chose only between $c_2$ and $c_4$, then one lacks 1 bit of free will. I am not claiming that this is how things work in reality, but I like to illustrate that if one takes free will on the one side and true randomness on the other side both seriously as real, then new interesting questions can be raised and answered.

It is well known that a proper demonstration of quantum nonlocality (i.e. violation of a Bell inequality) requires that the inputs are truly chosen at random, or even better by the free will of two independent humans, Alice and Bob. More precisely, if the inputs were predetermined, then no proper violation of any Bell inequality is possible, because the bound of the inequality could be as high as the algebraic maximum (the local variable could be correlated to the inputs or, equivalently, could determine which inputs are used in any run of the experiment). But what if the inputs are only partially chosen at random? For example, suppose a Bell inequality with four possible inputs on Alice side. What if Alice is actually determined in each round of the experiment to freely chose only between 2 of the 4 inputs (the 2 inputs may vary from one run to the next, so that on average all 4 inputs happen with equal frequency)? Would such a 1 bit lack of free will still allow for a violation of Bell inequality? And what if the input alphabet is arbitrarily large? Surprisingly, in \cite{BarrettGisinFreeWill} we prove that all probabilities of all standard (i.e. projective) measurements on two qubits in any maximally entangled state can be simulated if Alice lacks a single bit of free will, even if Bob enjoy full free will\footnote{Intuitively one may understand this by relating it to models based on the detection loophole. In one such model \cite{GisinG99} Bob always produces an outcome, but for any value of the local variable $\lambda$ Alice has only 50\% chances, depending on her input $x$, to produce an outcome. Now, if for all $\lambda$ Alice is programmed to chose only among those inputs that correspond to an outcome, then she always produces an outcome. This 50\% reduction of her free will corresponds to a one bit lack of free will.}. Consequently, if one of the two partners lacks a single bit of free will, then no violation of any Bell inequality with binary outcomes could ever be observed, whatever the number of possible inputs.

In the case of partially entangled pairs of qubits it is known that a lack of 2 bits of free will render any violation impossible \cite{BarrettGisinFreeWill}, but it is unknown whether the lack of a single bit of free will is compatible with a proper violation of a Bell inequality. For systems larger than qubits and for inequalities with more than binary outcomes the answers are unknown.

Is the above sketched connection between free will and Bell inequalities surprising? Maybe it isn't.
Nonlocal correlations seems to emerge, somehow, from outside space-time in the precise sense that no story in space-time can account for them \cite{SalartNature}.
Since free will also emerges somehow from outside space-time, Bell tests might provide us with glimpses on free will.

In summary, free will can be lower bounded (or the lack of free will upper bounded). Tests of Bell inequalities impose a lower bound on free will that is surprisingly low: physicists must enjoy much free will in order to perform meaningful Bell tests.

\section{Conclusion}\label{Concl}
Realistic true randomness, or in shorter and more elegant terms {\bf propensities}, allows one to consider new and interesting questions and applications. This is not surprising: scientists always made progress by considering the world around them as real and independent, with the possibility to interact with it. What is surprising is that so many good physicists interpret the violation of Bell's inequality as an argument against realism. Apparently their hope is to thus save locality, though I have no idea what locality of a non-real world could mean \cite{nonrealism}? It might be interesting to remember that no physicist before the advent of relativity interpreted the instantaneous action at a distance of Newton's gravity as a sign of non-realism (although Newton's nonlocality is even more radical than quantum nonlocality, as it allowed instantaneous signaling). Hence, it seems that the issue is not nonlocality, but non-determinism. In this note I argued that non-determinism and true randomness are not in contradiction with realism: propensities reflect pre-existing properties, but the reflection is not deterministic.
There is thus no conflict between realism and an open future: the future might not (yet) be real, but the process by which the future becomes actual is undoubtedly real.\\




\begin{thebibliography}{99}
\bibitem{propensitydef} I propose to use the word 'propensity' for the probability of a non-necessary (non-predetermined) physical event \cite{GisinPropensity}. Or as R. N. Giere writes \cite{Giere73}: "Propensities are . . . causal connections with varying strengths".
\end{thebibliography}
\end{document}